\documentclass[conference]{IEEEtran}
\IEEEoverridecommandlockouts
\usepackage{cite}
\usepackage{amsmath,amssymb,amsfonts}
\usepackage{listings}
\usepackage[linesnumbered, ruled, vlined]{algorithm2e}
\usepackage{graphicx}
\usepackage{textcomp}
\usepackage{xcolor}
\definecolor{codegreen}{rgb}{0,0.6,0}
\definecolor{codegray}{rgb}{0.5,0.5,0.5}
\definecolor{codepurple}{rgb}{0.58,0,0.82}
\definecolor{backcolour}{rgb}{0.95,0.95,0.92}
\lstdefinestyle{my_style}{
    backgroundcolor=\color{backcolour},   
    commentstyle=\color{codegreen},
    keywordstyle=\color{magenta},
    numberstyle=\tiny\color{codegray},
    stringstyle=\color{codepurple},
    basicstyle=\ttfamily\footnotesize,
    breakatwhitespace=false,         
    breaklines=true,                 
    captionpos=b,                    
    keepspaces=true,                 
    numbers=left,                    
    numbersep=5pt,                  
    showspaces=false,                
    showstringspaces=false,
    showtabs=false,                  
    tabsize=2
}
\def\BibTeX{{\rm B\kern-.05em{\sc i\kern-.025em b}\kern-.08em
    T\kern-.1667em\lower.7ex\hbox{E}\kern-.125emX}}
\begin{document}
\title{K-Detector: Identifying Duplicate Crash Failures in Large-Scale Software Delivery\\
}
\author{\IEEEauthorblockN{Hao Yang$^\ast$}
\IEEEauthorblockA{\textit{SAP Labs China} \\
Xi'an, China \\
ethan.yang@sap.com}
\and
\IEEEauthorblockN{Yang Xu$^\dagger$}
\IEEEauthorblockA{\textit{SAP Labs China} \\
Xi'an, China \\
yang.xu@sap.com}
\and
\IEEEauthorblockN{Yong Li$^\dagger$}
\IEEEauthorblockA{\textit{SAP Labs China} \\
Xi'an, China \\
yong.li@sap.com}
\and
\IEEEauthorblockN{Hyun-Deok Choi$^\dagger$}
\IEEEauthorblockA{\textit{SAP Labs Korea} \\
Seoul, South Korea \\
hyun.deok.choi@sap.com}
}
\maketitle
\begin{abstract}
After a developer submits code, corresponding test cases arise to ensure the quality of software delivery. Test failures would occur during this period, such as crash, error, and timeout. Since it takes time for developers to resolve them, many duplicate failures will happen during this period. In the delivery practice of SAP HANA, crash triage is considered as the most time-consuming task. If duplicate crash failures can be automatically identified, the degree of automation will be significantly enhanced. To find such duplicates, we propose a training-based mathematical model that utilizes component information of SAP HANA to achieve better crash similarity comparison. We implement our approach in a tool named Knowledge-based Detector (K-Detector), which is verified by 11,208 samples and performs 0.986 in AUC. Furthermore, we have deployed K-Detector to the production environment, and it can save 97\% human efforts in crash triage as statistics.
\end{abstract}
\section{Introduction}
Speed and efficiency are essential aspects of the software industry. In general, delivering a software product includes building, testing, releasing, and other processes. It is conceivable to automate software delivery, to make these practices transparent to the developer who has submitted code. Moreover, ``Quality deliveries with short cycle time need a high degree of automation.''~\cite{b1}, it intends to keep software products always in a release state, allowing the fast release of features and rapid responses to any failure. During the testing process, plenty of delays arise, such as crash, error, and timeout. Especially for a large-scale software system that contains several test suites for complex scenarios. SAP HANA is a database product that supports complex business analytical processes in combination with transactionally consistent operational workloads~\cite{b2}.

To our experience, about 35.2\% of delays are due to crash failures in the delivery practice of SAP HANA, and crash triage is considered as the most time-consuming task. It means developers cannot quickly get the feedback for the quality of their check-ins. Usually, when a crash failure occurs, a dump file containing stack frames will generate. Bettenburg et al. concluded that stack frame information is one of the most valuable items for developers~\cite{b3}. Schröter et al. explored bug fixing activities in Eclipse and found that it would be faster when a failing call stack is available~\cite{b4}. Stack traces provide developers insights about the place where a crash failure has occurred. However, stack traces can be rather verbose and require advanced domain knowledge and considerable time to investigate them. Therefore, we propose to leverage additional information within a project to enhance stack traces and reduce the required time investments to analyze crash failures.

Since it takes time for developers to resolve crash failures, many duplicates will happen during this period. If we can deal with the duplicate part in the early stage, it will save plenty of human efforts~\cite{b5, b6}. Jalbert et al. advised that the duplicate part should be ignored during the bug triage step ~\cite{b7}. Bettenburg et al. concluded that the additional information from the duplicate part could be utilized to achieve faster bug fixing by developers~\cite{b8}. Hence, it is useful and necessary to identify duplicate crash failures in software delivery.

In this paper, we implement our approach to identify duplicate crash failures through several steps. First, we process stack frames in a dump file to remove irrelevant information using regular expression extraction and stop word removal~\cite{b9}. Then, we convert the stack frame to a suitable abstraction level based on the knowledge we mined from the source-code repository. By selecting appropriate features and applying the training process, we can get the final formula to compare each call stack pair's similarity.

We evaluate our approach by using the crash data scrapped from Bugzilla~\cite{b10}, which is promising due to its AUC is 0.986. Furthermore, we implement our approach in a tool named Knowledge-based Detector (K-Detector) and apply it to the delivery practice of SAP HANA. It can save 97\% human efforts in crash triage as statistics.

This paper makes the following contributions:
\begin{itemize}
\item We convert the stack frame in crash dump file generated by SAP HANA into component information with more business meaning and use component sequence for crash similarity comparison.
\item We design a mathematical similarity model by selecting appropriate features and propose a training strategy to obtain better performance on metrics.
\item We apply our approach to the delivery practice of SAP HANA, which enhances the degree of automation and reduces human resource investment.
\end{itemize}
\begin{figure}[htbp]
\centerline{\includegraphics{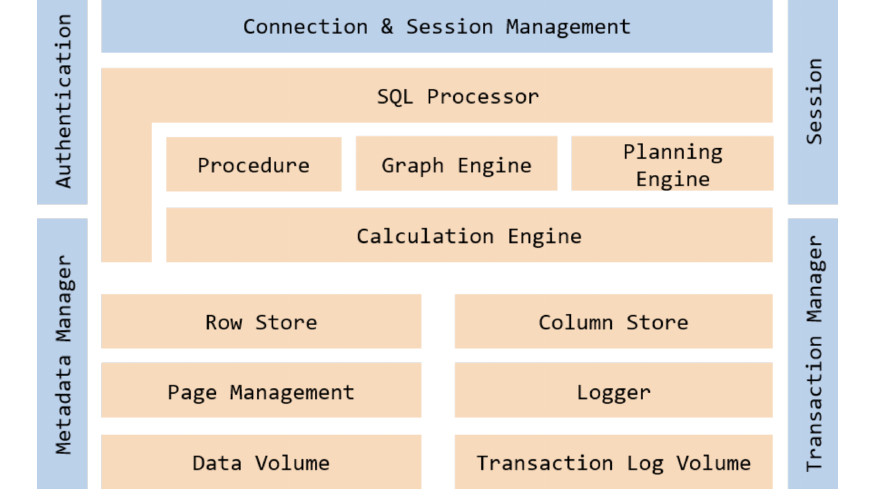}}
\caption{Architecture of Index Server.}
\label{fig1}
\end{figure}

The rest of this paper is organized as follows. Section~\ref{B} introduces the background information. We detail the implementation of each step separately in Section~\ref{TA}. Section~\ref{E} presents the experimental design and result. We discuss threats to validity in Section~\ref{TTV}. Section~\ref{RW} and Section~\ref{C} describe the related work and conclusion.
\section{Background}\label{B}
In this section, we introduce the background of SAP HANA architecture and crash dump structure.
\subsection{SAP HANA Architecture}
HANA is the core of SAP's data management platform~\cite{b11}. Using the columnar engine is a promising approach to deal with transactional and analytical workloads simultaneously. Furthermore, retrieval is the primary operation of OnLine Transactional Processing (OLTP) and OnLine Analytical Processing (OLAP), which will benefit from column-wise compression. Fig.~\ref{fig1} provides an overview of the architecture of SAP HANA Index Server, which has several components working together. It contains in-memory data stores and engines to process data.
\textbf{\begin{table}[htbp]
\caption{Number of functions/components in SAP HANA}
\begin{center}
\begin{tabular}{|c||c|c|c|c|}
\hline
{} & {2017} & {2018} & {2019} & {2020}$^{\mathrm{a}}$ \\
\hline
\# of functions & {371,873} & {416,223} & {369,108} & {405,838} \\
\hline
\# of components & {236} & {251} & {215} & {213} \\
\hline
\multicolumn{4}{l}{$^{\mathrm{a}}$Until May 31, 2020.}
\end{tabular}
\end{center}
\label{tab1}
\end{table}}

To compare the discrepancy between using components and functions, we measure the total number of their unique names in SAP HANA. Usually, the major components in a database perform significant tasks and contain plenty of functionalities. The result lists in Table~\ref{tab1}, the average number of functions is three orders of magnitude than the number of components. Hence, the use of components instead of functions is in a suitable abstraction level for identifying duplicate crash failures. We will show in Section~\ref{CWEM} that involving component information does not affect the practice in SAP HANA.
\subsection{Crash Dump Structure}\label{CDS}
\begin{figure}[htbp]
\centerline{\includegraphics{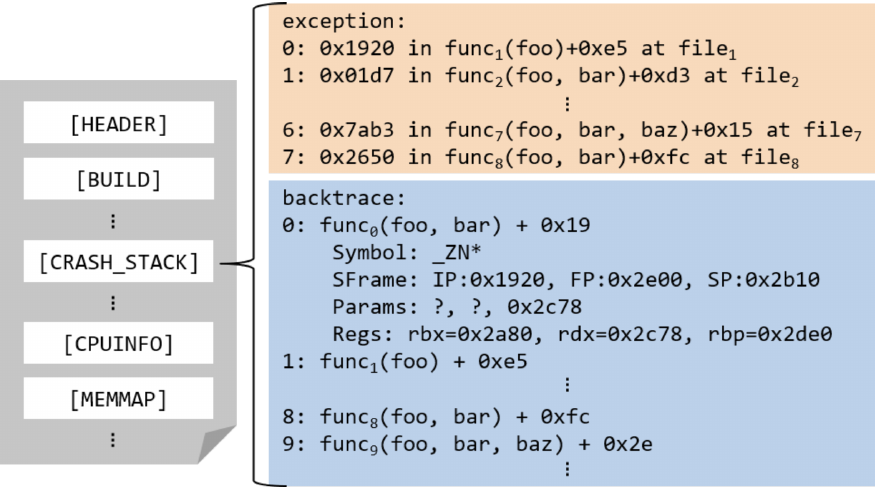}}
\caption{Structure of crash dump.}
\label{fig2}
\end{figure}

During the testing process of SAP HANA, some crash failures would occur with the dump file. It contains information about the software's state for developers to fix the bug, such as stack traces and register contents. Fig.~\ref{fig2} illustrates the structure of the crash dump file. At the top is [HEADER] that contains general information such as the Process ID (PID) of the crashing process and the time of crash failure.

We can see that there is a table of contents lists after [HEADER]. They are comprised of several sections such as [BUILD], [CRASH\_STACK], [CPUINFO], and [MEMMAP]. Furthermore, many HANA components have their sections with plenty of information. Since the content of the call stack exists in the [CRASH\_STACK], it is the part that we care about most. Sometimes call stack may contain an exception, and sometimes it is backtrace-only. The exception is the more crucial part of the call stack, and more likely to be the root cause of a crash failure.
\begin{figure}[htbp]
\centerline{\includegraphics{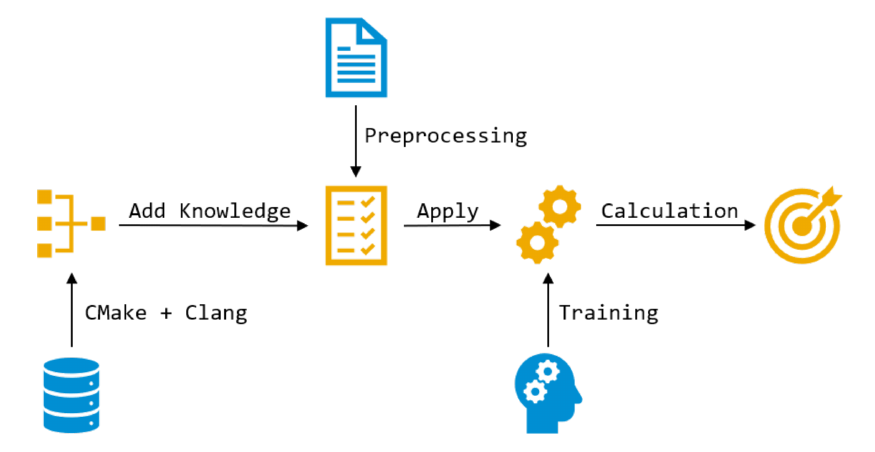}}
\caption{Workflow of our approach.}
\label{fig3}
\end{figure}
\section{The Approach}\label{TA}
\begin{figure}[htbp]
\centerline{\includegraphics{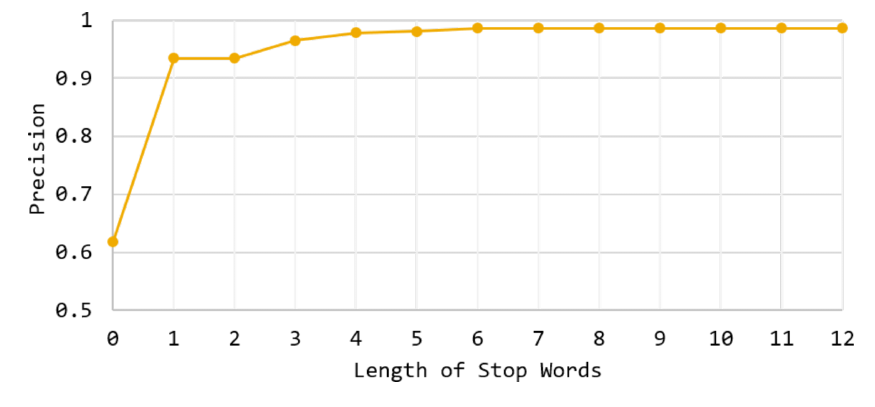}}
\caption{Enhancement of stop words.}
\label{fig4}
\end{figure}
This section details our approach for identifying duplicate crash failures of large-scale software delivery. Fig.~\ref{fig3} shows the overall workflow. The input is the source code, historical crash data, and a new crash dump file. The corresponding output is the duplicate crash dump that possibly hit. First, for each crash dump, we need to remove some irrelevant information through data preprocessing. Next, adding mined knowledge to convert the function-based call stack into a component sequence. Then, calculating each sequence pair's similarity through a designed mathematical model. Furthermore, the parameters of this model can learn from historical data.
\subsection{Data Preprocessing}\label{DP}
According to the explanations mentioned above, we first need to extract [CRASH\_STACK], as Fig.~\ref{fig2} shows. Although after preliminary data cleaning, the format is still too complex to compare. We need to remove irrelevant information, such as the offset address. Meanwhile, our approach only requires the function name that allows us to remove parameter variables and return type. The stack frame without a symbol is invalid since the linker uses it to link several parts of the program. Moreover, the detailed information can also be removed, including ``SFrame'', ``Params'', ``Regs'', etc.

As explained in Section~\ref{CDS}, the exception is the more crucial part and closer to the root cause. By comparing exception and backtrace, we can get some frequently occurring dispensable function names called stop words \cite{b9}. For the backtrace shown in Fig.~\ref{fig2}, the serial numbers of stop words are 0 and 9. Before calculating similarity, filtering stop words can effectively improve the precision, which is a key metric to evaluate our approach. Fig.~\ref{fig4} presents the precision enhancement by stop words, and the data comes from all crash dumps of SAP HANA in 2019. As the filtered stop words' length increases, we can see that the precision is enhanced significantly. When the curve is stable, we can obtain the stop words list for data preprocessing.
\subsection{Add Knowledge}\label{AK}
SAP HANA is developed by C++, using CMake to describe its building process. CMake is a meta build system that uses scripts called CMakeLists to generate build files for a specific environment \cite{b12}. An example of CMakeLists in SAP HANA is shown as follows.
\begin{lstlisting}[language=make]
# All files in this directory and its sub-
# directories belong to ComponentA
SET_COMPONENT("ComponentA")
# Except for File1 and File2 which belong to 
# ComponentB
SET_COMPONENT("ComponentB"
    File1
    File2
)
\end{lstlisting}
\begin{figure}[htbp]
\centerline{\includegraphics{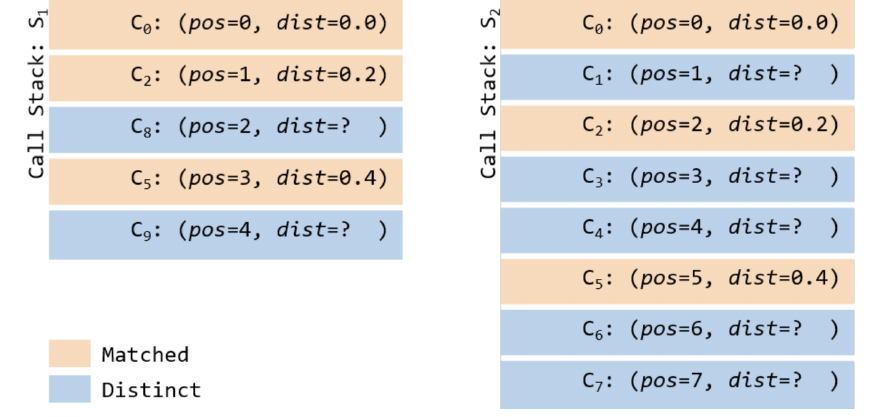}}
\caption{Illustration of call stack pair.}
\label{fig5}
\end{figure}

Directories or individual files can be associated with a specific component via the SET\_COMPONENT. The function takes either just a component name or a component name with a list of files. The former associates all files in the current directory with the specified component, and the later associates the specified files with the component. We use regular expression and Breadth-First Search (BFS) to extract Component-File mapping exits in the layered CMakeLists.

In computer science, lexical analysis is converting a sequence of characters into a token stream \cite{b13}. Then, syntactic analysis can take the output from lexical analysis and build a data structure called Abstract Syntax Tree (AST). We use the AST generated by Clang \cite{b14} to extract all fully qualified function names in the C++ file since Clang is light-weight and easy for secondary development. Finally, we can obtain File-Function mapping. With the relationship of Component-File that we have figured out, we can complete the conversion from function to a component.
\subsection{Mathematical Model}\label{MM}
After Section~\ref{DP} and Section~\ref{AK}, we have achieved the conversion from function-based call stack to a component sequence. Fig.~\ref{fig5} shows the form of such sequence, which is still a call stack but replaced with component information. The components in S$_1$ and S$_2$ are listed in order from top to bottom. We can see that the components in S$_1$ match the components in S$_2$ respectively (C$_0$, C$_2$ and C$_5$). It means that they have the same component name, and the similarity between call stack S$_1$ and S$_2$ is defined as follows.
\begin{equation}\label{S}
Similarity = \frac{\sum\nolimits_{cpnt\in LCS}e^{-m\times pos}e^{-n\times dist}}{\sum\nolimits_{i=0}^{max}e^{-m\times i}}
\end{equation}

In detail, $\sum_{cpnt\in LCS}$ means that we obtain Longest Common Subsequence (LCS) from each call stack pair and only accumulate this part. $pos$ represents the component position (from zero), each calculation takes the maximum position of matched components, and $m$ is a coefficient to adjust the rate of change caused by position. $dist$ stands for the component distance. It is a normalized edit distance \cite{b15} based on the function names in a component, and $n$ is a coefficient to adjust the rate of change caused by distance. $max$ represents the maximum possible component position in each call stack pair.

We choose the component position in a call stack and the component distance with multiple function names as two different features. As we know, a farther component from the top of a call stack should have a lower weight, and its influence will decrease by the increase of component position. Hence, we make use of $e^{-m\times pos}$ to describe component position feature. Likewise, a component that contains more similar function names should have a higher weight, and the influence should decrease as component distance increases. Hence, we use $e^{-n\times dist}$ to describe component distance feature.

Furthermore, to make the similarity result between 0 and 1, we need to divide numerator by a greater value. Hence, we use $\sum_{i=0}^{max}e^{-m\times i}$ as the denominator, and $m$ is the same as $m$ in the numerator. For a clearer understanding, through Equation~\ref{S}, the similarity calculation result of the call stack pair shown in Fig.~\ref{fig5} is 70.5\% (assuming $m=n=1$)\footnote{$Similarity=\frac{e^{-1\times 0}e^{-1\times 0.0}+e^{-1\times 2}e^{-1\times 0.2}+e^{-1\times 5}e^{-1\times 0.4}}{e^{-1\times 0}+e^{-1\times 1}+\cdots+e^{-1\times 7}}$}. Then, by comparing with the set threshold, we can conclude whether this call stack pair is duplicate.
\subsection{Training Strategy}
Our mathematical model uses two coefficients: $m$ is a coefficient related to the number of components that should be considered from the top of a call stack; $n$ is a coefficient associated with the level of component distance, which should be involved in similarity calculation. These parameter values can set randomly, but it will affect the result if they are not suitable. Hence, we propose a training strategy to do parameter tuning as follows.
\begin{algorithm}[htbp]
\caption{Parameter Tuning}
\KwIn{training set $D$;}  
\KwOut{optimal value pair $(m_{opt},\ n_{opt})$;}
$AUC_{max}\gets0$;\\
$(m_{opt},\ n_{opt})\gets(0,\ 0)$;\\
\For{$m=0.0;m\leq2.0;m\mathrel{+}=0.1$}
{
    \For{$n=0.0;n\leq2.0;n\mathrel{+}=0.1$}
    {
        Compute $AUC$ based on $D$; \\
        \If{$AUC>AUC_{max}$}
        {
            $AUC_{max}\gets AUC$;\\
            $(m_{opt},\ n_{opt})\gets(m,\ n)$;
        }
    }
}
return $(m_{opt},\ n_{opt})$;
\end{algorithm}

In this paper, we empirically choose the interval of $m$ and $n$ as [0.0, 2.0], and we increase $m$ and $n$ by 0.1 at each iteration. Next, the AUC calculation performs on all call stack pairs in the training set, and we record the value of $(m_{opt},\ n_{opt})$ when obtaining the optimal AUC. Finally, these tuned parameters return. The input of our tuning algorithm is a training set prepared in advance. It is based on the historical crash data in Bugzilla of SAP HANA. We sample positives and negatives by a crawling method, which will introduce in Section~\ref{CWEM}.
\begin{figure}[htbp]
\centerline{\includegraphics{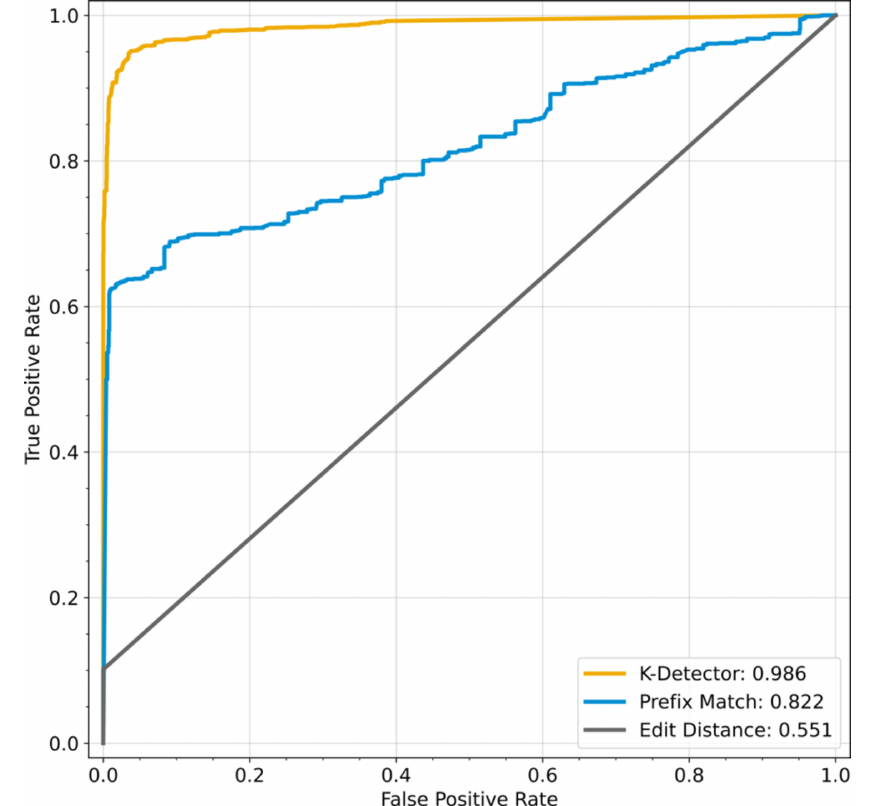}}
\caption{AUC comparison with existing methods.}
\label{fig6}
\end{figure}
\section{Evaluation}\label{E}
In this section, we conduct two kinds of evaluation. One compares our approach with existing methods, and another is collecting feedback after our approach has been deployed to the production environment.
\subsection{Compare with Existing Methods}\label{CWEM}
In the delivery practice of SAP HANA, test failures are tracked by Bugzilla. We choose crash failure as the target for our approach and have designed a crawling method. All duplicate crash dumps\footnote{In 2019, about 34.49\% of crash failures are duplicate.} can be crawled automatically through RESTful APIs such as ``resolution'', ``creation\_time'', and ``cf\_crashdump\_location''. Then, applying the Union-Find algorithm \cite{b16} on these dump files and use ``dupe\_of'' attribute as input, we can obtain the group information of crash dumps. Next, we do sampling, the crash dump pair of the same group forms a positive sample through combination. We select the dump pair of different groups, but from the same component, to get a negative sample. After that, we can obtain 5,604 positive samples and 5,604 negative samples (derived from 2019). They are divided into two sets for training and testing.

\begin{figure}[htbp]
\centerline{\includegraphics{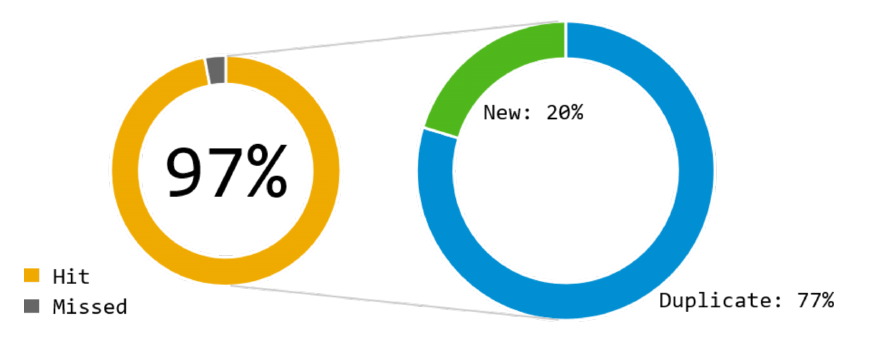}}
\caption{Feedback collection from delivery practice.}
\label{fig7}
\end{figure}
At the current time, there are many similarity calculation methods for call stack matching. Each of these matching methods assigns to a call stack pair a similarity score ranging between 0 (for distinct) and 1 (for identical). Edit Distance \cite{b15} is a popular measure for string similarity, and it is also the matching method previously used in SAP HANA. It defines the edit distance of call stack pair as the number of edit operations (add, delete, replace, etc.) needed to convert the first call stack into the second call stack on a character level. Prefix Match is based on the fundamental premise that two call stacks arising from the same problem will have the same function names closer to the top \cite{b17}.

We compare our approach with these two existing methods through AUC based on the testing set. Fig.~\ref{fig6} shows that the prefix match performs better than the previously applied method, but our approach has a more obvious enhancement.
\subsection{Feedback from Delivery Practice}
Our approach has been deployed to the production environment as K-Detector for identifying duplicate crash failures in the delivery practice of SAP HANA. For the crash failure that just occurred in the testing process, if our tool detects duplicates in recent failures, it will automatically bind the related Bug ID to this failure. Besides, if K-Detector does not find a duplicate crash failure, a new bug will be auto-created for developers to track this problem promptly. Hence, the delivery pipeline will not be affected by such delay, i.e., crash failure.

We collected feedback from the engineering team, which contains 203 recent crash failures. It is a different dataset from Section~\ref{CWEM}, and the results present in Fig.~\ref{fig7}. We can see that 97\% of crash failures (197) are \textit{Hit}, which means that K-Detector can achieve the same effect as manual operations. Besides, the remaining 3\% of crash failures (6) are \textit{Missed}, and we will discuss this part in detail in Section~\ref{FP}. 

Furthermore, in the \textit{Hit} part, there are 77\% of crash failures (157) K-Detector can find duplicates, which will reduce the number of bug filing and improve the efficiency of bug fixing. Besides, we can see that 20\% of crash failures (40) are new crash failures that did not occur before. Although our tool cannot detect duplicates in such case, new bugs are auto-filed immediately to track these problems. As described above, K-Detector can save 97\% human efforts in crash triage.
\section{Threats to Validity}\label{TTV}
This section discusses the validity and generalizability of our approach. In particular, we discuss false positives, applied scenarios, and timely updates.
\subsection{False Positives}\label{FP}
False positives will mislead our approach into binding a wrong Bug ID to current crash failure. Compared with false negatives, whose defect is to create more bugs, false positives will cause more damage to our approach. From Section~\ref{E}, we can see that our approach's false positive rate is not zero, and K-Detector also has some \textit{Missed} cases. It reveals that sometimes the call stack pair shares some irrelevant components at the top, and these components have a great weight according to our approach.
\subsection{Applied Scenarios}
Although our approach is based on the component information of SAP HANA, there are similar concepts, such as package and module, in other large-scale software implemented by object-oriented language. It is conceivable to utilize our approach to analyze other target systems. Due to discrepancies in architecture and business logic, we probably obtain a different result and need to make slight adjustments when applying our approach.
\subsection{Timely Updates}
Function-Component mapping is the term we use to cope with call stack conversion. Since this mapping relationship usually changes with the code changes, it will incur inaccurate conversion if the update is not timely. Moreover, the improper update of trained parameters will also affect crash similarity comparison.
\section{Related Work}\label{RW}
In recent years, there are several works based on different approaches for similar bug matching in a software system. We classify them into Information Retrieval (IR) and Machine Learning (ML) to discuss respectively.
\subsection{Information Retrieval}
IR aims to extract useful information from unstructured documents, which are expressed by natural language. Runeson et al. proposed an approach using Natural Language Processing (NLP) to detect duplicate defect reports, which processes the words from plain English and uses word occurrence statistics to identify similar reports \cite{b6}. Then, Saha et al. introduced an approach named BLUiR to improve bug localization, which extracts the model and code constructs, such as class and method names as structured documents \cite{b18}. Next, Wang and Lo proposed an approach named AmaLgam to continue improving bug localization, which puts version history, similar reports, and structure together \cite{b19}. Then, Ye et al. introduced an adaptive approach to rank relevant files for a bug report, which uses domain knowledge such as API descriptions to bridge the lexical gap between bug reports and source code \cite{b20}. These approaches rely on submitted bug reports and focus on solving developers' problems when fixing bugs. However, our approach applies to the delivery pipeline, and the purpose is to improve software delivery efficiency. Moreover, we provide a training strategy to cover more changes made to a software system.
\subsection{Machine Learning}
ML aims to select some features from the bug report and design a way to measure the report pair's similarity based on these features. Sabor et al. proposed a feature extraction approach named DURFEX for detecting duplicate bug reports efficiently, which starts by abstracting call stack into a sequence of defined packages \cite{b21}. It is similar to our approach that converts function-based call stack into a component sequence to achieve better crash similarity comparison. Nevertheless, DURFEX keeps track of the call stack sequence by creating feature vectors of n-grams with varying. It does not innovate the distance between the package and the top of a call stack, which may affect the detection result. Besides, our approach has a parameter related to the number of components that should be considered from the top.

Deshmukh et al. introduced an approach using Siamese Convolutional Neural Network (SCNN) and Long Short-Term Memory (LSTM) to detect and retrieve duplicate and similar bugs accurately \cite{b22}. It seeks to leverage deep learning advances for natural language processing and separately handles short and long descriptions using different neural network architectures. However, the approach relies on the artificial description of bugs, and K-Detector can identify duplicates with the original crash dump file, thereby reducing extra manual operations.
\section{Conclusion}\label{C}
In this paper, we propose an approach to detect duplicate crash failures in software delivery. We observe how developers use their knowledge to find duplicate crash failures, and then we attempt to simulate and automate these processes. By judging whether the crash failure is duplicate, the delivery pipeline can automatically handle crash delays, saving much time. We convert the function-based call stack to a component sequence through the knowledge we mined from the source-code repository. Then, we design a mathematical model using component position and component distance to achieve better crash similarity comparison. Through data evaluation, our approach has a significant enhancement over existing methods, whose AUC is 0.986. Furthermore, in the delivery practice of SAP HANA, our approach also has an outstanding achievement, which can save 97\% human efforts in crash triage.

In the future, we plan to classify the mined component information of SAP HANA and innovate ``stop components'' reference to the idea of ``stop words'' to reduce the false positives threat. We will do some experiments on other large-scale software systems to verify our approach's feasibility in different applied scenarios. Moreover, we plan to develop a service to do knowledge mining and model training periodically to reduce the update delay.
\section*{Acknowledgment}
We want to express our sincere thanks to Dong-Won Hwang and Sascha Schwedes from SAP Labs for their engineering support; Prof. Yoo Shin from KAIST for his academic support, which makes K-Detector a reality. We also thank the anonymous reviewers for their feedback.


\begin{thebibliography}{00}
\bibitem{b1} C. Ebert, G. Gallardo, J. Hernantes and N. Serrano, ``Devops'', IEEE Software, vol. 33, no. 3, pp. 94-100, 2016.
\bibitem{b2} F. Färber, S. K. Cha, J. Primsch, C. Bornhövd, S. Sigg and W. Lehner, ``SAP HANA database: Data management for modern business applications'', ACM SIGMOD Rec., vol. 40, no. 4, pp. 45-51, 2011.
\bibitem{b3} N. Bettenburg, S. Just, A. Schröter, C. Weiss, R. Premraj and T. Zimmermann, ``What makes a good bug report?'', Proceedings of the 16th ACM SIGSOFT International Symposium on Foundations of software engineeing, pp. 308-318, 2008.
\bibitem{b4} A. Schröter, N. Bettenburg and R. Premraj, ``Do stack traces help developers fix bugs?'', Proc. MSR, pp. 118-121, 2010.
\bibitem{b5} J. Lerch and M. Mezini, ``Finding duplicates of your yet unwritten bug report'', Proc. of European Conference on Software Maintenance and Reengineering, pp. 69-78, 2013.
\bibitem{b6} P. Runeson, M. Alexandersson and O. Nyholm, ``Detection of duplicate defect reports using natural language processing'', Proceedings of the 29th international conference on Software Engineering, pp. 499-510, 2007.
\bibitem{b7} N. Jalbert and W. Weimer, ``Automated duplicate detection for bug tracking systems'', Proc. International Conference on Dependable Systems and Networks With FTCS and DCC, pp. 52-61, 2008.
\bibitem{b8} N. Bettenburg, R. Premraj, T. Zimmermann and S. Kim, ``Duplicate bug reports considered harmful... really?'', Proc of International Conference on Software Maintenance, pp. 337-345, 2008.
\bibitem{b9} G. Lohman, J. Champlin and P. Sohn, ``Quickly finding known software problems via automated symptom matching'', Proceedings of the Second International Conference on Automatic Computing, pp. 101-110, 2005.
\bibitem{b10} N. Serrano and I. Ciordia, ``Bugzilla, itracker and other bug trackers'', IEEE Software, vol. 22, no. 2, pp. 11-13, 2005.
\bibitem{b11} V. Sikka, F. Färber and W. Lehner, ``Efficient transaction processing in SAP HANA database: The end of a column store myth'', Proc. ACM SIGMOD Int. Conf. Manag. Data, pp. 731-742, 2012.
\bibitem{b12} K. Martin and B. Hoffman, ``Mastering CMake: A Cross-Platform Build System'', Clifton Park, NY, USA:Kitware, 2010.
\bibitem{b13} K. Louden, ``Compiler Construction: Principles and Practice'', Boston, USA: PWS Publishing, 1997.
\bibitem{b14} C. Lattner, ``LLVM and Clang: Next generation compiler technology'', in The BSD Conference, 2008.
\bibitem{b15} V. I. Levenshtein, ``Binary codes capable of correcting deletions insertions and reversals'', Doklady Akademii Nauk SSSR, vol. 163, no. 4, pp. 845-848, 1965.
\bibitem{b16} B. A. Galler and M. J. Fisher, ``An improved equivalence algorithm'', Commun. ACM, vol. 7, no. 5, pp. 301-303, 1964.
\bibitem{b17} N. Modani, R. Gupta, G. Lohman, T. Syeda-Mahmood and L. Mignet, ``Automatically identifying known software problems'', In ICDE Workshops, pp. 433-441, 2007.
\bibitem{b18} R. K. Saha, M. Lease, S. Kunshid and D. E. Perry, ``Improving bug localization using structured information retrieval'', Proc. Int. Conf. Autom. Softw. Eng., pp. 345-355, 2013.
\bibitem{b19} S. Wang and D. Lo, ``Version history similar report and structure: Putting them together for improved bug localization'', ICPC'14, pp. 53-63, 2014.
\bibitem{b20} X. Ye, R. Bunescu and C. Liu, ``Learning to rank relevant files for bug reports using domain knowledge'', FSE'14, pp. 689-699, 2014.
\bibitem{b21} K. Sabor, A. Hamou-Lhadj and A. Larsson, ``DURFEX: A feature extraction technique for efficient detection of duplicate bug report'', Proc. of the IEEE International Conference on Software Quality Reliability and Security, pp. 240-250, 2017.
\bibitem{b22} J. Deshmukh, A. K. M, S. Printder, S. Sengupta and N. Dubash, ``Towards accurate duplicate bug retrieval using deep learning techniques'', 2017 IEEE International Conference on Software Maintenance and Evolution (ICSME), pp. 115-124, 2017.
\end{thebibliography}
\end{document}